\documentclass[preprint]{aastex}

\slugcomment{Apr.\ 02, 2000}

\def\dfrac#1#2{{\displaystyle\frac{#1}{#2}}}

\def\sd{{\rm d}}
\def\sg{{\rm g}}
\def\oxy{{\rm O}}

\def\sG{{\rm G}}
\def\SFR{{\rm SFR}}
\def\chisq0{{\chi_0}^2}
\def\med{{\rm median}}
\def\psiro{\Psi_{{\rm obs},i}^{\rm raw}}
\def\psico{\Psi_{{\rm obs},i}^{\rm cor}}
\def\psiim{\Psi_i^{\rm int}}
\def\psim{\Psi_i}
\def\xif{X_i^{\rm f}}

\begin{document}

\title{TESTING INTERMITTENCY OF THE GALACTIC STAR FORMATION HISTORY
ALONG WITH THE INFALL MODEL}

\author{\bf Tsutomu T. Takeuchi\altaffilmark{1} and
Hiroyuki Hirashita\altaffilmark{1}}
\affil{Department of Astronomy, Faculty of Science, Kyoto University,
Sakyo-ku, Kyoto 606-8502, Japan}
\affil{E-mail: takeuchi, hirasita@kusastro.kyoto-u.ac.jp}

\altaffiltext{1}{Research Fellows of the Japan Society for the Promotion of
Science.}

\begin{abstract}

We analyze the star formation history (SFH) of the Galactic disk 
by using an infall model. Based on the observed SFH of the Galactic
disk, we first determine the timescales of the gas infall
into the Galactic disk ($t_{\rm in}$) and that of the gas
consumption to form stars ($t_{\rm sf}$). Since each of the two
timescales does not prove to be determined independently from the
SFH, we first fix $t_{\rm sf}$. Then, $t_{\rm in}$ is determined so that
we minimize $\chi^2$. Consequently, we choose three parameter
sets: $(t_{\rm sf}\,{\rm [Gyr]},\, t_{\rm in}\,{\rm [Gyr]})=
(6.0,\, 23)$, (11, 12), and  (15, 9.0), where we set the
Galactic age as 15 Gyr. All of the three cases
predict almost identical star formation history.
Next, we test the intermittency (or variability) of the
star formation rate (SFR) along with the smooth
SFH suggested from the infall model. The large value of the $\chi^2$ 
statistic supports the violent time-variation of the SFH.
If we interpret the observed SFH with smooth and variable
components, the amplitude of the variable component is comparable
to the smooth component.
Thus, intermittent SFH of the Galactic disk is strongly suggested.
We also examined the metallicity distribution of G-dwarfs.
We found that the true parameter set lies between $(t_{\rm sf}\,{\rm [Gyr]},
\, t_{\rm in}\,{\rm [Gyr]})= (6, 23) \mbox{ and }(11, 12) $, though
we should need a more sophisticated model including the process of metal
enrichment within the Galactic halo.

\end{abstract}


\section{INTRODUCTION}\label{sec:intro}

In the field of the Galactic chemical evolution,
\citet{Eggen62} have inspired the modeling.
From the correlation between the ultraviolet excess and orbital
eccentricity of stars, they concluded that the Galaxy formed
by collapse on a free-fall timescale from a single protogalactic
cloud. An alternative picture of halo formation is proposed
by \citet{Searle78}. They argued that Galactic system is
formed from the capture of fragments such as dwarf galaxies
over a longer timescale than that proposed by Eggen et al.
In any case, determining the timescale of the infall of matter and
the chemical enrichment is an important problem to resolve 
the formation mechanism of the Galactic
system. 
Indeed, there have been a number of papers that investigated
the formation and chemical evolution of the Galaxy
(e.g., \citealt{Matteucci89}) and spiral galaxies (e.g.,
\citealt{LB75}; \citealt{Sommer96}).

Many models of the star formation history (SFH) of the Galaxy
includes the formation of the Galactic disk by gas infall. This
scenario (so-called the infall model) is consistent with the
age-metallicity relation of the disk stars
(e.g., \citealt{Twarog80}), if a reasonable SFH is applied. Moreover, 
the infall model provides
a physically reasonable way of solving the G-dwarf problem
(e.g., \citealt{Pagel97}, p.236), contrary to the
closed-box model which tends to overpredict the number of the
low-metallicity stars.

One of the factors that determine the star formation rate (SFR)
is the gas content of galaxies. 
Indeed, the SFR and the gas density is closely
related to each other \citep{Kennicutt98}, and this relation is
generally called the Schmidt law \citep{Schmidt59}.
The SFR and the gas density, $\rho$, are related as $\SFR\propto\rho^n$,
where $n=1 - 2$.
In the one-zone model, the relation is simply assumed to be
described with
$\SFR\propto M_{\rm g}^n$, where $M_{\rm g}$ is the gas mass
in the galaxy.

Though the ``classical'' infall model is widely accepted,
there are observational data that suggest the intermittent star
formation activities in spiral galaxies.
Kennicutt, Tamblyn, \& Congdon (1994) showed the ratio of present-to-past SFR
in spiral sample has a significant scatter. More recently,
Tomita, Tomita \& Sait\={o} (1996) analyzed the far-infrared to $B$-band 
flux ratio $f_{\rm FIR}/f_B$ of 1681 spiral galaxies
(see also \citealt{Devereux97}).
The indicator $f_{\rm FIR}/f_B$ also represents the ratio between the
present SFR and the averaged SFR during the recent Gyr.
They showed the order-of-magnitude spread of $f_{\rm FIR}/f_B$ and suggested
the violent temporal variation of the SFR.

The intermittency of the SFH in the Galactic disk is recently suggested 
by Rocha-Pinto et al.\ (2000; hereafter R00).
They derived the SFH of the Galactic disk from observed
age distribution of the late-type stars and suggested
that the star formation activity of the disk is intermittent or 
violently variable.
They tested to show that the variation is of statistical significance. 
Their test are based on the null hypothesis that the SFR of the Galaxy is
temporally ``constant'', but their derived SFH indicates that the SFR 
around 10--15 Gyr ago is significantly smaller than that around 0--10 Gyr ago. 
This trend of the Galactic SFR can be interpreted along with the infall 
scenario of the Galactic disk formation:
Gas-infall to the Galactic disk occurred in the first
$\sim$ 5 Gyr. 
Thus it is an important work to interpret their data
through the infall model and to examine whether the variation of
SFR is still significant even in the infall scenario.

In this paper, we re-analyze the data by R00 in the context
of the infall model. In the next section, we briefly
review the derivation of the SFH in the Galactic disk by R00. 
Next, in \S\ref{sec:infall}, we present the formulation
of the infall model, based on which we test the data by R00 statistically. 
In \S\ref{sec:discuss}, we interpret the observed
SFH by two components; underlying infall-model component and 
violently variable component.
In the same section, possible physical mechanisms for the
variation of star formation are also presented.
Finally, we summarize the content in \S\ref{sec:summary}.

\section{THE DATA}

R00 provided the SFH of the Galaxy inferred from the stellar age of
the solar neighborhood.  
They used 552 late-type stars.
The age of each star is estimated from the chromospheric
emission in the $\mbox{Ca}\,${\sc ii} H and K lines \citep{Soderblom91}. 
After metallicity-dependent age correction, completeness correction, 
and scale-height correction\footnote{The scale height is dependent on 
the age of stars.}, they derived the age distribution of the stars. 
Then, after correcting the stars that is not alive at present, 
they derived the SFH of the Galactic disk in their Figure~2. 
The discussion in this paper is based on Figure~2 of R00.

R00 concluded that the inferred SFH is representative of the
SFH of the whole disk, since the timescale of the diffusion of stars
to the kpc scale is $\sim 0.2$ Gyr, which is much shorter than the
Galactic age. 
Thus, in the next section, we model the Galactic
SFH by a one-zone model to extract the {\it global} property of
the SFH. The difference between the one-zone and
the multi-zone treatment is seen in Figure~6 of \citet{Sommer96},
where we see that the one-zone treatment is a good approximation.

\section{SFH FROM INFALL MODEL}\label{sec:infall}

In this section, we give a physical interpretation of the
SFH by R00. First, we adopt the
infall model for the interpretation, since the model has been
successful in reproducing the age-metallicity relation
of stars in the solar neighborhood (e.g., \citealt{Pagel97}).
Next, we infer the timescales of the following two processes;
the infall of gas into the Galactic disk and the gas consumption
to form stars. Finally, we examine whether the intermittent
SFH proposed in R00 is of statistical significance.

\subsection{Model Description}\label{subsec:model}

We assume the one-zone model: The gas mass in the Galaxy
is treated as a function of time, $M_\sg (t)$. 
The time evolution of the gas mass is determined by the infall
from the halo, whose rate is described by $F(t)$, 
the consumption by star formation, $\psi (t)$, and the
recycling from stellar mass loss. 
If the instantaneous recycling approximation is adopted \citep{Tinsley80}, 
the time evolution of the gas mass in the Galaxy is described by
\begin{eqnarray}
  \frac{\sd M_{\sg}}{\sd \, t} = - (1 - R) \psi + F \; ,
\label{basic_gas} 
\end{eqnarray}
where $R$ is the returned fraction from stellar mass loss, described by
\begin{eqnarray}
  R = \int_{m_t}^{m_u}(m-w_m) \phi (m) \,\sd m\; .
\label{recycle}
\end{eqnarray}
Here, $\phi (m)$ is the initial mass function (IMF) of stars,
$m_t$ is the present turnoff mass ($1M_\odot$), $m_u$ is the upper
cutoff of the stellar mass ($100M_\odot$), and $w_m$ is the remnant
mass (we assume $w_m=0.7M_\odot$ for $m<4M_\odot$ and $w_m=1.4M_\odot$
for $m>4M_\odot$).
The IMF is normalized so that the integral of $m\phi (m)$ in
the full range of the stellar mass becomes unity.
When the Salpeter IMF, $\phi (m)\propto m^{-2.35}$ (\citealt{Salpeter55}) 
with the lower cutoff of $0.1 \, M_\odot$ and the upper cutoff of
$100 \, M_\odot$ is assumed, $R=0.32$.
\citet{Pagel97} derived a similar value for $R$ by using
a different form of IMF and remnant mass ($R=0.2$ -- $0.3$),
and commented that the uncertainty in $R$ is $\sim 0.1$.

Since the normalization of the SFR and statistical test become much more 
complicated unless the SFR is presented by analytic function, we adopt 
the instantaneous recycling approximation for clarity in this section.
The analytic form makes it significantly easy to perform the statistical 
test in \S\ref{subsec:statistics}.
We examine the propriety of adopting the instantaneous recycling 
approximation to this problem in \S\ref{subsec:non_ira}.

In this paper, the timescale of the gas infall onto the Galactic disk,
$t_{\rm in}$, will be determined by a fitting to the observational data 
in \S\ref{subsec:statistics}.
For the convenience of the fitting, we assume the infall rate is expressed by 
an analytic function.
A natural form is the following exponential function (\citealt{Pagel97}, 
p.242):
\begin{eqnarray}
  F(t)=\frac{M_0}{t_{\rm in}}\exp (-t/t_{\rm in}),\label{infall}
\end{eqnarray}
where $M_0$ indicates the total mass that can fall into disk; in other words,
\begin{eqnarray}
  \int_0^\infty F(t)\,\sd t=M_0.
\end{eqnarray}
Normalizing the equation (\ref{basic_gas}) by $M_0$ leads
\begin{eqnarray}
  \frac{\sd f_{\sg}}{\sd \, t} = - (1 - R) \frac{f_\sg}{t_{\rm sf}}  
  + \frac{1}{t_{\rm in}}\exp\, (-t/t_{\rm in}) \; , \label{modified_gas}
\end{eqnarray}
where
\begin{eqnarray}
f_\sg \equiv \dfrac{M_\sg}{M_0} \; .
\end{eqnarray}
The Schmidt law of $n=1$ (\S\ref{sec:intro}) is assumed as follows:
\begin{eqnarray}
  \tilde{\psi}\equiv \psi/M_0 = f_\sg /t_{\rm sf} \; ,\label{def_tilde}
\end{eqnarray}
where $t_{\rm sf}$ indicates the timescale for the gas to be
converted to stars (i.e., the gas consumption timescale).
Then, we can express the solution of equation (\ref{modified_gas}) 
analytically as
\begin{eqnarray}
  f_\sg (t) &=& \dfrac{\beta}{1 - \beta} \left( e^{-t/t_{\rm in}} - 
    e^{-(1-R)t/t_{\rm sf}}\right) \; ,\label{gas_fraction} \\
  \beta &\equiv& \frac{t_{\rm sf}}{(1 - R) t_{\rm in}} \; .
\end{eqnarray}
One of the goals in this paper is to infer the two timescales,
$t_{\rm in}$ and $t_{\rm sf}$ from the SFH in R00.
Thus, we define the SFR at the Galactic age of $t$, $\tilde{\psi}$, as 
\begin{eqnarray}
  \tilde{\psi} = \tilde{\psi} (t \,; \, t_{\rm in}, t_{\rm sf})\equiv 
  f_{\rm g}/t_{\rm sf}\; .\label{sfr_parameter}
\end{eqnarray}

For the following discussions it is convenient to define the averaged
SFR for the normalization of $\psi$. 
The averaged SFR, $\langle\tilde{\psi} \rangle$, is defined by
\begin{eqnarray}
  \langle \tilde{\psi} \rangle\equiv \frac{1}{T_\sG} \int_{0}^{T_\sG}
  \tilde{\psi} \,\sd t .
\end{eqnarray}
Equations (\ref{gas_fraction}) and (\ref{sfr_parameter}) lead
the averaged SFR as
\begin{eqnarray}
  \langle\tilde{\psi} \rangle = \frac{1}{T_\sG} \int_{0}^{T_\sG} 
  \dfrac{\beta }{(1 - \beta )t_{\rm sf}}
  \, \left( e^{-t/t_{\rm in}} - e^{-(1-R)t/t_{\rm sf}}
  \right) \, \sd t,
\end{eqnarray}
where $T_\sG$ is the age of the Galactic disk and we take $T_\sG = 15$ Gyr 
following R00.
Then we obtain
\begin{eqnarray}
  \dfrac{\tilde{\psi}}{\langle\tilde{\psi} \rangle} 
  = \dfrac{ e^{-\tau /\tau_{\rm in}} - 
    e^{-(1-R)\tau /\tau_{\rm sf}} }
  { \tau_{\rm in} \left( 1 - e^{-1/\tau_{\rm in}}\right) 
    -\dfrac{\tau_{\rm sf}}{(1-R)} \,
    \left( 1 - e^{-(1-R) /\tau_{\rm sf}} \right)
    } \equiv 
  \Psi (\tau \,; \, \tau_{\rm in}, \tau_{\rm sf})\; ,
  \label{Psi}
\end{eqnarray}
where $\tau\equiv t/T_\sG$, $\tau_{\rm in}\equiv t_{\rm in}/T_\sG$, and
$\tau_{\rm sf}\equiv t_{\rm sf}/T_\sG$.
Here we note that 
\begin{eqnarray}
  \int_{0}^{1} \Psi (\tau )\, \sd \tau = 1\; .\label{normalization}
\end{eqnarray}

\subsection{Trend Estimation}

\begin{figure}[t]
\figurenum{1}
\centering\includegraphics[width=10cm]{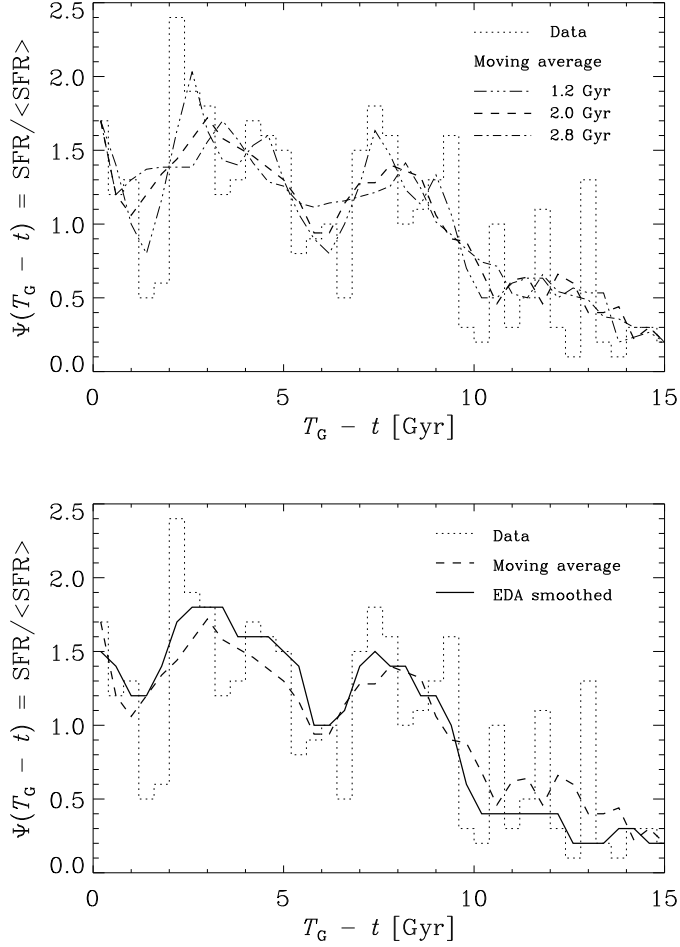}
\figcaption[fig1.ps]{
The smoothed star formation history of the Galaxy. 
In the upper panel, we show the smoothing result by moving average.
The dot-dot-dot-dashed line is the smoothed SFR with smoothing kernel 
width of 1.2 Gyr, strong dashed line is the smoothed SFR 
with 2.0 Gyr kernel, and dot-dashed line is the smoothed SFR with 2.8 Gyr 
kernel.
In the lower panel, the dashed line represents the moving average of 
2.0 Gyr kernel, and the solid line is the smoothed SFH by the moving 
median method of the exploratory data analysis.
We observe that the 1.2-Gyr kernel is not sufficient to smooth the 
varying SFH.
}\label{fig1} 
\end{figure}

We extract the overall ``trend'' from the data of R00.
We used the smoothing method developed in the field of the exploratory 
data analysis (EDA), as well as the ordinary moving average.
The EDA smoothing is based on the moving median, which is is known to 
be quite robust against outliers in the datasets compared with the 
moving average (e.g., Hoaglin, Mosteller, \& Tukey 1983). 
Since the SFH in R00 violently varies with time, the EDA procedure
is expected to be suitable for the problem.
Detailed procedures are summarized in the appendix.
The smoothed results with the above two procedures are shown in
Figure~\ref{fig1}.
The original data are depicted by the dotted histogram.
In the upper panel, we show the smoothing results by moving average.
Details of the moving average method is extensively discussed in e.g., 
\citet{Hart97}.
The dot-dot-dot-dashed line is the smoothed SFR with smoothing kernel 
width of 1.2 Gyr, strong dashed line is the smoothed SFR 
with 2.0 Gyr kernel, and dot-dashed line is the smoothed SFR with 2.8 Gyr 
kernel.
In the lower panel, the dashed line represents the moving average of 
2.0 Gyr kernel, and the solid line is the EDA smoothed SFH.
We observe that the 1.2-Gyr kernel is not sufficient to smooth the 
varying SFH.
This means that the timescale of the variation is less than $\sim 2$ Gyr. 
Possible mechanisms of the variation that works in less than
2 Gyr are listed in \S\ref{subsec:mechanism}.

We see that the two results yield consistent trends, but the EDA result
is not affected by the sudden jump of the values compared with the moving 
average as clearly seen at $T_{\rm G}-t=10$ -- 15 Gyr.
We find the rise of the SFR in the early epoch. 
This is interpreted as being the gradual increase of gas by the infall.
Thus we suggest that the overall SFH of the Galaxy is well described by 
the infall scenario. 
We statistically infer the infall timescale and the gas consumption 
timescale in the next section. 
We will also examine whether the `residual' deviated from the trend
is of statistical significance there.

\subsection{Parameter Estimation and Statistical Test}
\label{subsec:statistics}

In this subsection we perform a statistical test for the observed 
Galactic (normalized) SFR, $\Psi_{\rm obs}$ (shown as
$\SFR /\langle\SFR\rangle$ in R00) with respect to that 
inferred from the infall model.
The observed data are binned as ${\Psi_{\rm obs}}_i$, $i=1, 2, \dots, k$.
Once the parameter set $(\tau_{\rm in},\,\tau_{\rm sf})$
is fixed, we can estimate these parameters from the modified Pearson's 
chi-square statistic, $\chisq0$ (see e.g., \citealt{Rao73}, p.352):
\begin{eqnarray}
  \chisq0 &\equiv& \sum_{i=1}^k \dfrac{(n_i - N\psiim)^2}{n_i} 
  = N \sum_{i=1}^k \dfrac{(\psiro - \psiim)^2}{\psiro}\; ,
\end{eqnarray}
where $n_i$ is the raw number of stars in the $i$-th bin, 
$N$ is the sample size (in this case $N = 552$),
$\psiro$ is the raw normalized SFR without completeness corrections, and
$\psiim$ is the binned theoretical SFR which we would have observed under
the same condition as $\psiro$.
However, since the SFR in Figure~2 of \citet{Rocha00a} is corrected for 
the selection biases when they have derived the value, we must include
the effect of the incompleteness correction in the statistical analysis.
We set $n_{{\rm cor},i} = c_i n_i$, then we obtain
\begin{eqnarray}
  c_i = \dfrac{n_{{\rm cor},i}}{n_i} = 
  \dfrac{N_{\rm cor}\psico}{N_{\rm raw}\psiro} = 
  \dfrac{\psico}{\psiro} \;,
\end{eqnarray}
where $\psico$ is the corrected SFR as shown in R00.
Here, we fix the data size as $N_{\rm cor} = N_{\rm raw} = N$.
As long as $N_{\rm cor} \sim N_{\rm raw}$, the
statistical significance is not affected by this procedure.
Therefore we have the model value
\begin{eqnarray}
  \psim = c_i \psiim \; ,
\end{eqnarray}
where
\begin{eqnarray}
  \Psi_i\equiv\frac{1}{\tau_{i+1}-\tau_i}
  \int_{\tau_i}^{\tau_{i+1}}\Psi (\tau )\,\sd\tau \;.
\end{eqnarray}
In addition, R00 note that the error bar is a Poisson 
fluctuation, thus the $i$-th error bar, $\sigma_i$, can be described as
\begin{eqnarray}
  \sigma_i = \frac{c_i}{N} \left( N\psiro \right)^{1/2} 
  = \left( \dfrac{c_i\psico}{N}\right)^{1/2}\; .
\end{eqnarray}
Considering the above, we observe
\begin{eqnarray}
  {\chi_0}^2 &=& N \sum_{i=1}^k \dfrac{
          \left(\dfrac{\psico}{c_i} - \dfrac{\psim}{c_i} \right)^2}{
          \dfrac{\psico}{c_i}}
        = N \sum_{i=1}^k \dfrac{(\psico - \psim)^2}{c_i\psico}\nonumber \\
        &=& \sum_{i=1}^k \dfrac{(\psico - \psim)^2}{{\sigma_i}^2} \; .
\end{eqnarray}
For the numerical convenience,
we fix $\psico =\Psi_i=0$ at $\tau =0$, and
the other 37 points are used for the inference (i.e., $k=37$).

\begin{figure}[t]
\figurenum{2}
\centering\includegraphics[width=10cm]{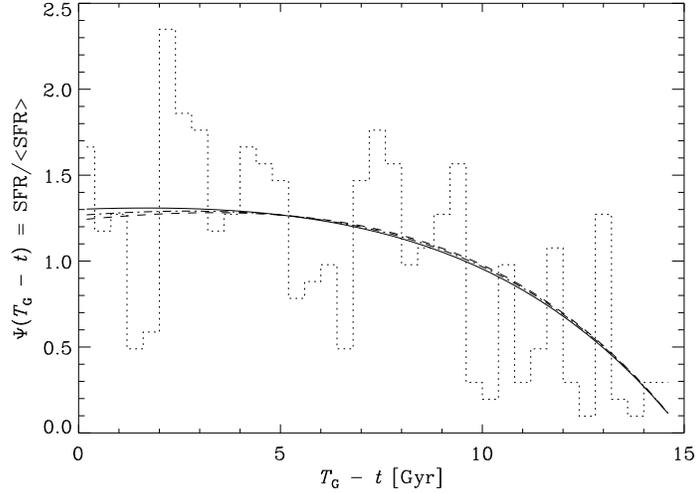}
\figcaption[fig2.ps]{Star formation histories for the
parameter sets in Table \ref{tab1}. The dotted line is the
observed star formation history in R00. The solid, dashed,
and dash-dotted lines present the cases of
$(\tau_{\rm sf},\,\tau_{\rm in})=(0.4,\, 1.5),$ (0.7, 0.8), and
(1.0, 0.6), respectively, though it is difficult to distinguish
the three lines.
\label{fig2}}
\end{figure}

As we will see later, $\tau_{\rm sf}$ and $\tau_{\rm in}$
are not determined independently. 
Thus, we will fix one of the parameters.
Since $\tau_{\rm sf}$ is extensively investigated by
\citet{Kennicutt94}, we first fix $\tau_{\rm sf}$. Then,
$\tau_{\rm in}$ is determined by minimizing $\chisq0$. 
As representative values for the gas consumption timescale, $\tau_{\rm sf}$,
we choose 0.4, 0.7 and 1.0 ($t_{\rm sf}=6.0$, 11, and 15 Gyr,
respectively)\footnote{If we set $\tau_{\rm sf}<0.4$,
we obtain an unreasonably large $\tau_{\rm in}(\ga 2)$ for the best fit
parameter.}. 
The best-fit $\tau_{\rm in}$ and $\chisq0$ for each $\tau_{\rm sf}$ 
are listed in Table \ref{tab1}.
The best-fit $\tau_{\rm in}$'s are 1.5, 0.8, 0.6, respectively,
and $\chisq0 =172$.
The three cases are presented in Figure~\ref{fig2}. We see that
the three parameter sets describes an almost identical SFH.
This means that the two parameters, $\tau_{\rm sf}$ and $\tau_{\rm in}$
are strongly correlated and it is almost impossible to determine 
$\tau_{\rm sf}$ and $\tau_{\rm in}$ independently.

We are able to test the goodness of fit between $\Psi_{\rm obs}$ and 
$\Psi_i$ at the same time by evaluating $\chisq0$ with respect to the 
$\chi^2$-statistics with $(k - 2)$ degrees of freedom.
If $\chisq0 > \chi^2 (k - 2, \alpha)$, then the hypothesis that the 
observed SFH is produced by the infall scenario is rejected 
with confidence level $(1 - \alpha)$. 
In fact, even if we set $\alpha =0.01$, $\chi^2 (k - 2, \alpha)=60$. 
Thus, we conclude that the data of R00 is not produced by 
the ``classical'' infall model.
This clearly indicates the fact that the Galactic SFH is not continuous, but
strongly intermittent or variable.

\section{DISCUSSIONS}\label{sec:discuss}

\subsection{Metallicity and G-dwarf Problem}

One of the prime motivation for infall models is
that they provide a physically reasonable way of solving
the G-dwarf problem (\citealt{Pagel97} and references therein).
Thus, in this section, we examine the chemical evolution
of the Galaxy.

Under the one-zone treatment and the instantaneous recycling
approximation, the time evolution of the abundance of
the heavy element, whose species is labeled by $i$ ($i={\oxy }$,
C, Si, Mg, Fe, ...),
is expressed as
\begin{eqnarray}
\frac{\sd M_i}{\sd t}=-X_i\psi +E_i+\xif F(t),\label{basic_metal}
\end{eqnarray}
where $M_i$, $X_i$, $E_i$ and $\xif$ are the total mass of
heavy element $i$, the abundance of $i$ (i.e., $X_i\equiv M_i/M_{\rm g}$),
the total injection rate of element $i$ from stars, and
the abundance of the infalling gas, respectively \citep{Tinsley80}. 
Here, $E_i$ is expressed as
\begin{eqnarray}
  E_i=(RX_i+Y_i)\psi ,\label{ira_metal}
\end{eqnarray}
where $R$ is defined in equation (\ref{recycle}) and
$Y_i$ is the mass fraction of the element $i$ newly produced
and ejected by stars; in other words,
\begin{eqnarray}
  Y_i = \int_{m_t}^{m_u} mp_i(m)\phi (m)\,\sd m,
\end{eqnarray}
where $p_i(m)$ is the fraction of mass converted into the element
$i$ in a star of mass $m$. 
Adopting the Salpeter's IMF and the stellar yield by \citet{maeder92},
$Y_{\oxy }=1.8\times 10^{-2}$, where we adopt the oxygen
as a tracer of heavy elements (i.e., $i={\oxy }$;
\citealt{lisenfeld98,Hirashita99}). Combining equations
(\ref{basic_gas}), (\ref{infall}), (\ref{def_tilde}),
(\ref{basic_metal}), and (\ref{ira_metal}), we obtain
\begin{eqnarray}
  f_\sg\frac{\sd X_i}{\sd\tau}=\frac{f_\sg Y_i}{\tau_{\rm sf}}-
  \frac{X_i-\xif}{\tau_{\rm in}}e^{-\tau /\tau_{\rm in}}.
\label{basic_metal2}
\end{eqnarray}
Since the time-evolution of $f_\sg$ is solved in equation
(\ref{gas_fraction}), equation (\ref{basic_metal2}) can be integrated
to obtain $X_i$ as a function of $\tau (=t/T_{\rm G})$. 
We hereafter choose oxygen as a tracer of the metallicity; i.e., $i={\oxy }$
and assume $X_\oxy^{\rm f}=0.01X_{\oxy\, ,\odot}$.
For a more detailed modeling, we should include the metal
enrichment within the Galactic halo (e.g., \citealt{Ikuta99}).

\begin{figure}[t]
\figurenum{3}
\centering\includegraphics[width=10cm]{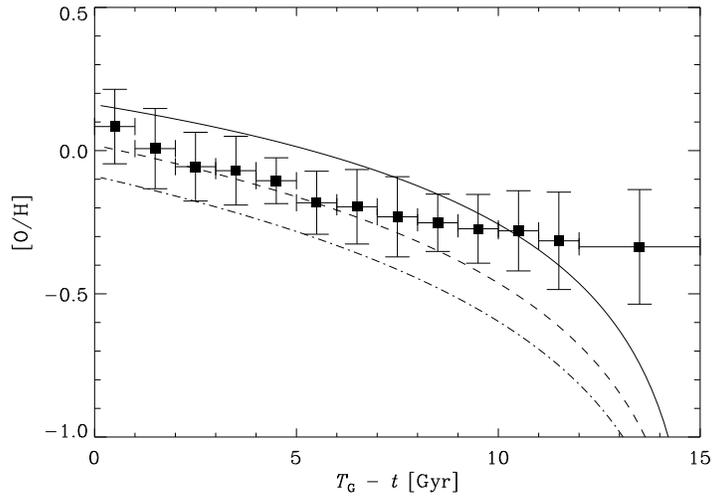}
\figcaption[fig3.ps]{Time evolution of oxygen abundance
$X_{\oxy}$ for the three sets of the
parameters in Table \ref{tab1}. The solid, dashed,
and dash-dotted lines present the cases of
$(\tau_{\rm sf},\,\tau_{\rm in})=(0.4,\, 1.5),$ (0.7, 0.8), and
(1.0, 0.6), respectively. Contrary to Fig.~\ref{fig2},
we can easily distinguish the three lines.
The true parameter set seems to lie between $(\tau_{\rm sf},\,
\tau_{\rm in}) =(0.4, 1.5)$ (the solid line) and $(\tau_{\rm sf},\,
\tau_{\rm in})= (0.7, 0.8)$ (the dashed line), though uncertainty and 
discrepancy exist.
\label{fig3}}
\end{figure}

The result is shown in Figure \ref{fig3} for the three parameter sets in
Table \ref{tab1}, where $X_{{\oxy },\,\odot}=0.013$ represents the
solar oxygen abundance (\citealt{Whittet92}, p.42).
The solid, dashed,
and dash-dotted lines present the cases of
$(\tau_{\rm sf},\,\tau_{\rm in})=(0.4, 1.5), (0.7, 0.8), \mbox{ and }
(1.0, 0.6)$, respectively.
We also present the observational data of age-metallicity relation
by \citet{Rocha00b} (see their Table 3), where we assume that
${\rm 0.7\,[Fe/H]=[O/H]} \equiv \log (X_{\oxy }/X_{{\oxy },\,\odot})$
by using \citet{Edvardsson93}.
The true parameter set which reproduce the observational 
data points seems to lie between the solid line $(\tau_{\rm sf},\, 
\tau_{\rm in}) = (0.4,1.5)$ and $(\tau_{\rm sf},\,\tau_{\rm in}) = 
(0.7, 0.8)$, though the discrepancy at the lower-metallicity side is 
prominent.
But if the infalling gas is more enriched than have been assumed, 
this discrepancy might be resolved.
We should also note the large scatter of the relation (Fig.~13
of \citealt{Rocha00b}) and the uncertainty in the above relation
between [Fe/H] and [O/H].

The infall timescale is larger than that in \citet{Sommer93}, 
who gave 3.4 Gyr. 
The difference comes not only from the different way of modeling but also
from the no prominent decline of the recent SFR as presented
in Figure~\ref{fig2}.

\begin{figure}[t]
\figurenum{4}
\centering\includegraphics[width=10cm]{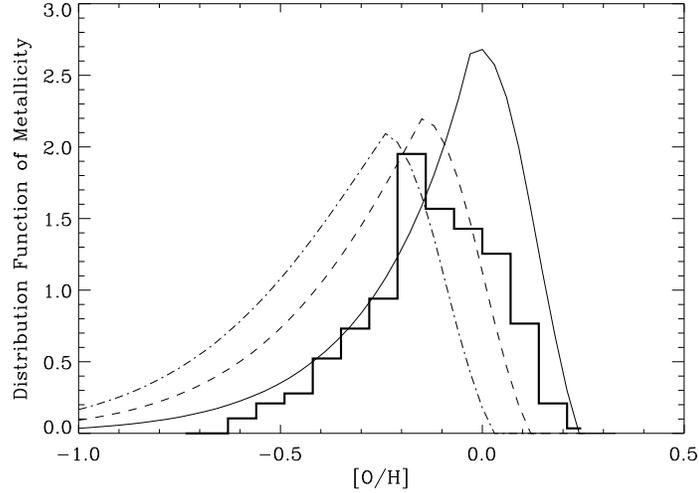}
\figcaption[fig4.ps]{Distribution of the G-dwarf metallicity.
for the three sets of the
parameters in Table \ref{tab1}. The solid, dashed,
and dash-dotted lines present the cases of
$(\tau_{\rm sf},\,\tau_{\rm in})=(0.4, 1.5), (0.7, 0.8), \mbox{ and }
(1.0, 0.6)$, respectively. 
The histogram shows the data by Rocha-Pinto \& Maciel (1996). 
Again the true parameter set seems to lie between $(\tau_{\rm sf},\,
\tau_{\rm in}) =(0.4, 1.5)$ (the solid line) and $(\tau_{\rm sf},\,
\tau_{\rm in})= (0.7, 0.8)$ (the dashed line), though uncertainty and 
discrepancy exist.
\label{fig4}}
\end{figure}

The G-dwarf problem is also tested along with our model.
The probability distribution function $P(\log X_{\oxy })$ of the
metallicity is calculated from our model as
\begin{eqnarray}
  P(\log X_{\oxy })\,\sd\log X_{\oxy }=\Psi\,\sd\tau .\label{prob_X}
\end{eqnarray}
Using equations (\ref{basic_metal2}) and (\ref{prob_X}), we obtain
the following analytical expression for $P$:
\begin{eqnarray}
  P(\log X_{\oxy}) = (\ln 10)\; \Psi (\tau )X_{\oxy }(\tau )\left[
    \frac{Y_{\oxy }}{\tau_{\rm sf}}-
    \frac{X_{\oxy}(\tau )-X_\oxy^{\rm f}}
    {\tau_{\rm in}f_\sg (\tau )}e^{-\tau /\tau_{\rm in}}
  \right]^{-1} \;,
\end{eqnarray}
where $\tau$ is a function of $X_{\oxy }$ and the functional form
is determined by solving equation (\ref{basic_metal2}).
In comparing the distribution function with the observational
data, we should take into account the scatter of the data. 
Here, we simply convolve $P$ with a Gaussian kernel as
\begin{eqnarray}
  P_{\rm conv}(\log X_\oxy )\equiv\int_{-\infty}^\infty P(u)\,
  \frac{1}{\sqrt{2\pi}\,\sigma}\,
  \exp\left[ -\frac{(\log X_\oxy -u)^2}{2\sigma^2}\right]\,\sd u,
\end{eqnarray}
where we adopt $\sigma =0.1$ to compare with \citet{Rocha96}.
We adopt these data because we would like to use a sample of
G-dwarfs, whose lifetime is as long as the age of the universe.
The result is shown in Figure~\ref{fig4}.
The solid, dashed, and dash-dotted lines present the cases of
$(\tau_{\rm sf},\,\tau_{\rm in}) = (0.4,1.5), (0.7, 0.8), \mbox{ and }
(1.0, 0.6)$, respectively. 
The histogram shows the data by \citet{Rocha96}, where 
${\rm 0.7\,[Fe/H]=[O/H]}$ is assumed and the integrated number 
is normalized to unity. 
We see the same trend as Figure \ref{fig3}:
Again we see that the true parameters lie between the solid line 
and the dashed line.
As mentioned above, we should include the detailed enrichment
process within the Galactic halo as e.g., in \citet{Ikuta99}.
However, since the aim of this paper is to examine the
intermittent SFH proposed by R00, we do not examine the
chemical evolution further.

\subsection{Comment on the Instantaneous Recycling Treatment}
\label{subsec:non_ira}

The instantaneous recycling approximation is adopted
only because it provides a very convenient way to obtain an
analytic SFH, which is easily applied to the statistical analysis
in \S\ref{subsec:statistics}. 
However, if we fix $t_{\rm sf}$, the instantaneous recycling treatment
tends to overestimate the rate of gas consumption to form stars due to the
instantaneously recycled gas, which should be recycled with
a delay in the realistic situation. 
Thus, we here examine the effect of the delayed recycling. 
We examine the opposite extreme case where the effect of the delayed 
recycling is significant: The gas is never recycled.
In other words, we examine the case of $R=0$.

\begin{figure}[t]
\figurenum{5}
\centering\includegraphics[width=10cm]{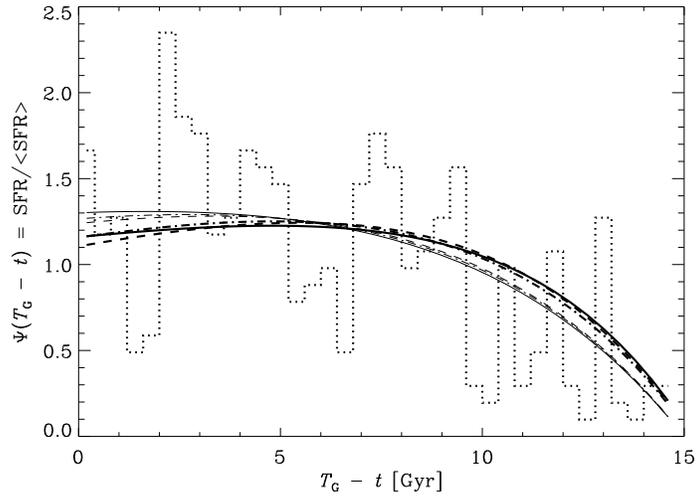}
\figcaption[fig5.ps]{Star Formation History in the non-recycling
model ($R=0$) for the three parameters ({\it thick lines}$\,$) as an
extreme case, where the timescale of the recycling is very large. 
The star formation history in the instantaneous recycling model is 
also shown ({\it thin lines}).
The solid, dashed, and dash-dotted lines present the cases of
$(\tau_{\rm sf},\,\tau_{\rm in})=(0.4, 1.5), (0.7, 0.8), \mbox{ and }
(1.0, 0.6)$ , respectively.
\label{fig5}}
\end{figure}

In Figure~\ref{fig5}, we present the result for $R=0$ for the three
parameters $(\tau_{\rm sf},\,\tau_{\rm in})=(0.4,\, 1.5),$ (0.7, 0.8),
and (1.0, 0.6) with the thick lines. 
The thin lines represent the case where $R=0.32$; i.e., the same as
Figure~\ref{fig3}. 
We adopt the same normalization for both sets of lines 
(equation~\ref{normalization}). 
Comparing the thick and thin lines, we see the more decline of the SFR in 
the recent epoch in the non-recycling case than the thin lines.
This is because the gas consumption timescale is shorter in the 
non-recycling case \citep{Kennicutt94}. 
However, we stress that the variation of the observed
SFH (R00) is so large that the difference between the
instantaneous and non-instantaneous treatments does not
affect the conclusion that the SFR is violently variable.

\subsection{Two-Component Model of the Star Formation Rate}

\begin{figure}[t]
\figurenum{6}
\centering\includegraphics[width=10cm]{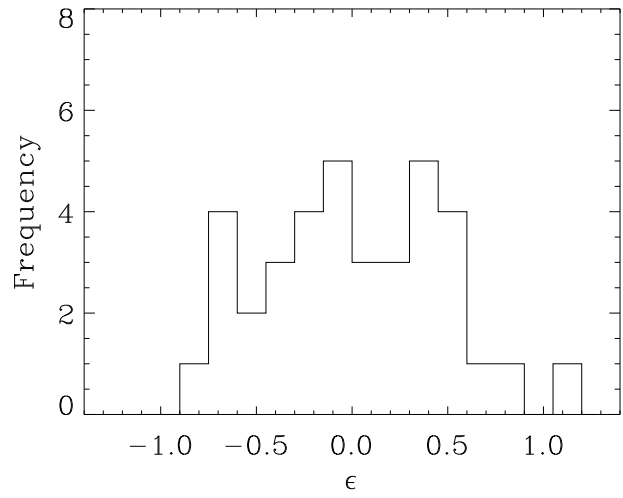}
\figcaption[fig6.ps]{Distribution of the residual component of the star
formation rate, $\varepsilon$. 
\label{fig6}}
\end{figure}

Since the observed SFH cannot be reproduced by a simple infall model,
we introduce a stochastically varying residual and modify the model 
as follows:
\begin{eqnarray}
    \Psi_{\rm obs} (\tau) = \Psi_{\rm infall}(\tau) + \varepsilon \; ,
\end{eqnarray}
where $\Psi_{\rm infall} (t)$ is the estimated best-fit infall model in the 
above discussion, and the residual, $\varepsilon$, is produced by a 
probability distribution with zero mean and dispersion $\sigma^2$.
We simply assume that $\sigma^2$ is time-independent.
Since $\varepsilon$ is estimated for each of the 37 bins,
we can show a distribution of $\varepsilon$ (Figure~\ref{fig6}).
The estimated sample dispersion $\sigma^2 = 0.22$ ($\sigma = 0.47$).
Considering the uncertainty in the age estimation in R00, 
the time-variation of the SFR is blurred with the uncertainty, as commented 
in R00. 
Thus, the above value of $\sigma$ can still be underestimated. 
Therefore, we conclude that the $\sigma$ is at least comparable to 
(perhaps larger than) $\Psi_{\rm infall}$.

If we assume that the large variation
of the SFR is typical of spiral galaxies, the star formation
activities of them should show a variety.
The large value of the variance ($\sigma$) is consistent with
previous works that suggested the variety of star formation
activities of spiral galaxies \citep{Kennicutt94,Tomita96,Devereux97}. 
Furthermore, the kurtosis of the residual, $K = -0.81$, which means
that $\varepsilon$ is distributed flatly and is not strongly 
concentrated around the mean. 
Indeed, in the Figure~8 of \citealt{Tomita96},
there seems to be little concentration of star
formation activity around the mean, which is consistent with the
flat distribution in Figure~\ref{fig6}.

\subsection{Possible Mechanisms for the Variation of SFH}
\label{subsec:mechanism}

Here, we consider what mechanisms are possible for the violent
time-variation of SFR in the Galactic disk. 
We mention the following two mechanisms.

The first possibility is that the infall into the Galactic disk is
a stochastic process. 
If an infalling gas is in the form of a cloud or a small-sized galaxy, 
such gas may induce a burst of star formation and increase SFR 
instantaneously. 
Indeed, infall of small galaxy seems to occur frequently seeing that the 
Sagittarius dwarf galaxy is now infalling to the Galaxy \citep{Ibata94}. 
The high-velocity clouds \citep{Wakker97} may fall into the Galactic disk 
stochastically and induce stochastic bursts of star formation.

The second possibility is that the interstellar medium is
a non-linear open system. 
A non-linear open system often shows a limit-cycle behavior of 
physical quantities \citep{Nicolis77}. 
The application of the non-linear-open-system model to the interstellar 
medium is described in \citet{Ikeuchi83}. 
According to their model, the fractional mass of the cold component 
($X_{\rm c}$) can oscillatory change. 
Recently, \citet{Kamaya97} suggested that the SFR varies oscillatory 
if the Schmidt law of $\SFR\propto X_{\rm c}^2$ is assumed, where
$X_{\rm c}$ is the mass ratio of the cold gas to the whole
gas (see also \citealt{Ikeuchi88} for a review and
\citealt{Hirashita00} for a recent development on this theme).

\section{SUMMARY}\label{sec:summary}

In the context of the infall model, we re-analyzed the SFH of
the Galaxy derived observationally by R00. 
We test to examine whether variation of the star formation rate 
proposed by R00 is significant.

We first statistically infer the timescales of the gas infall
into the Galactic disk ($t_{\rm in}$) and that of the gas
consumption to form stars ($t_{\rm sf}$). 
Since each of two timescales are not determined independently, 
we first fix $t_{\rm sf}$. 
Then, $t_{\rm in}$ is determined so that we minimize $\chi^2$. 
Consequently, we choose three parameter sets: 
$(t_{\rm sf}\,{\rm [Gyr]},\, t_{\rm in}\,{\rm [Gyr]})=
(4.5, 33), (10.5, 9.0), \mbox{ and } (15, 7.5)$, where we set the
Galactic age as 15 Gyr. 
All of the three cases predict an almost identical SFH. 
The parameter set that seems to fit best to the age-metallicity relation
and the metallicity distribution of the G-dwarfs is 
$(t_{\rm sf}\,{\rm [Gyr]},\, t_{\rm in}\,{\rm [Gyr]})= (10.5,\, 9.0)$. 
The infall timescale is larger than that in \citet{Sommer93}, 
who gave 3.4 Gyr. 
The difference comes not only from the different way of modeling but also
from the no prominent decline of the recent SFR as presented 
in Figure~\ref{fig2}.

Next, we test the intermittency (or violent variability) along
with the smooth SFH suggested from the infall model. 
The large value of $\chi^2$ statistic supports the violent variation of 
the SFH. 
Then, we interpret the observed SFH with the two components; underlying
smooth component described by the infall model, and violently
variable component. 
We find that the variation of the latter component is comparable to
the former.

As a test of the models, we also examine the age-metallicity relation
and the metallicity distribution of the Galactic stars.
Consequently, we observe that the degeneracy of the three parameters
are resolved and find that the true parameter set seems to lie between 
$(t_{\rm sf}\,{\rm [Gyr]},\, t_{\rm in}\,{\rm [Gyr]})= (10.5,\, 9.0) 
\mbox{ and } (4.5, 33)$.

Finally, two physical mechanisms for the variable SFH are suggested. 
One is the stochastic infall of the clouds or small galaxies, 
and the other is the non-linear oscillation of the SFR due to the 
limit-cycle behavior of the fractional mass of the cold-gas component.

\acknowledgments

We first thank J.~Sommer-Larsen, the referee, whose comments
much improved the quality of this paper.
We also thank H.~Kamaya, A.~Tomita, T.~T.~Ishii, S.~Mineshige, and M.~Sait\={o}
for useful discussions and kind helps. 
We are grateful to
K.~Yoshikawa for an excellent computational environment.
This work is supported by the Research Fellowship of the Japan
Society for the Promotion of Science for Young Scientists.
We made extensive use of the NASA's Astrophysics Data System Abstract 
Service (ADS).

\appendix
\section{SMOOTHING WITH MOVING MEDIAN}

In this appendix, we explain the details of the EDA smoothing based on the 
moving median.
The procedure is as follows:
\begin{enumerate}
\item We take the moving median of three sequential values of the original 
        data. 
        For the boundary data, we use the median of the following:
\begin{eqnarray}
        \hat{x}_{1} &=& \med (x_2 - 2(x_2 - x_3), x_1, x_2)\; , \\
        \hat{x}_{n} &=& \med (x_n + 2(x_{n-1} - x_{n-2}), x_{n-1}, x_n)\; ,
\end{eqnarray}
        where $\hat{x}_i$ is the smoothed value.
\item We perform the splitting of the plateau of the median-smoothed data 
        sequence, i.e. when we find the same value $\hat{x}_i$ and 
        $\hat{x}_{i+1}$, we substitute 
\begin{eqnarray}
        \hat{\hat{x}}_i &=& \med (x_{i-1} + 2(x_{i-1}-x_{i-2}), 
        x_i, x_{i-1}), \\
        \hat{\hat{x}}_{i+1} &=& \med (x_{i+2} - 2(x_{i+3}-x_{i+2}), 
        x_{i+1}, x_{i+2}).
\end{eqnarray}
into $\hat{x}_i$ and $\hat{x}_{i+1}$, respectively.
\item We, then, take the mean of $\hat{x}_{i-1}$ and $\hat{x}_{i+1}$ 
        (we denote them as $\tilde{x}_i$). 
        Finally, we derive the mean of $\hat{x}_i$ and  $\tilde{x}_i$.
\end{enumerate}
This method is often used in the econometric and biometric researches,
and provides successful results.


\newpage

\begin{table}[htb]
 \caption{Examined Parameters and $\chisq0$.}

 \begin{tabular}{ccccc}
  \hline\hline
   $\tau_{\rm sf}$\tablenotemark{a} & $t_{\rm sf}$ (Gyr) & 
   $\tau_{\rm in}$\tablenotemark{b} & $t_{\rm in}$ (Gyr) &
   $\chisq0$ \\ \hline
   0.4 & 6.0 & 1.5 & 23 & 172 \\
   0.7 & 11  & 0.8 & 12 & 172 \\
   1.0 & 15  & 0.6 &  9 & 172 \\
 \hline
 \end{tabular}
 \tablenotetext{a}{The gas consumption timescale ($t_{\rm sf}$)
 normalized by $T_\sG =15$ Gyr.}
 \tablenotetext{b}{The gas infall timescale ($t_{\rm in}$) normalized 
 by $T_\sG =15$ Gyr.}
\label{tab1}
\end{table}

\clearpage

\end{document}